\documentclass[final,aps,PRB,10pt, twocolumn, superscriptaddress, longbibliography, nobalancelastpage]{revtex4-2}
\usepackage{graphicx}
\usepackage{graphics}
\usepackage{amsmath}
\usepackage{amssymb}
\usepackage{amsfonts}
\usepackage{dsfont}
\usepackage{braket}
\usepackage[usenames,dvipsnames]{xcolor}
\usepackage{physics}
\definecolor{darkblue}{rgb}{0, 0, 0.8}
\usepackage[colorlinks=true, breaklinks=true, linkcolor=red, citecolor=blue, urlcolor=blue]{hyperref} 
\usepackage{hyperref}
\usepackage{subfigure}
\usepackage{xfrac}
\usepackage{bm}
\usepackage{kantlipsum}
\usepackage{enumitem}
\usepackage{tikz}
\usepackage{framed}
\usepackage{graphicx}
\usepackage{subfigure}
\usepackage{booktabs}

\begin{document}

\title{Dynamics of Negativity of a Wannier-Stark Many-Body Localized System Coupled to a Bath}


\author{Elisabeth Wybo} 
\affiliation{Department of Physics, Technical University of Munich, 85748 Garching, Germany}
\affiliation{Munich Center for Quantum Science and Technology (MCQST), Schellingstr. 4, 80799 M\"unchen}
\author{Michael Knap} 
\affiliation{Department of Physics, Technical University of Munich, 85748 Garching, Germany}
\affiliation{Institute for Advanced Study, Technical University of Munich, 85748 Garching, Germany}
\affiliation{Munich Center for Quantum Science and Technology (MCQST), Schellingstr. 4, 80799 M\"unchen}
\author{Frank Pollmann} 
\affiliation{Department of Physics, Technical University of Munich, 85748 Garching, Germany}
\affiliation{Munich Center for Quantum Science and Technology (MCQST), Schellingstr. 4, 80799 M\"unchen}



\begin{abstract}
An interacting system subjected to a strong linear potential can host a many-body localized (MBL) phase when being slightly perturbed. This so-called Wannier-Stark or `tilted-field' MBL phase inherits many properties from the well-investigated disordered MBL phase, and provides an alternative route to experimentally engineer interacting localized systems without quenched disorder. In this work, we investigate the dynamics of entanglement in a Wannier-Stark MBL system coupled to a dephasing environment. As an accessible entanglement proxy, we use the third R\'{e}nyi negativity $R_3$, which reduces to the third R\'{e}nyi entropy in case the system is isolated from the environment. This measure captures the characteristic logarithmic growth of interacting localized phases in the intermediate-time regime, where the effects of the coupling to the environment are not yet dominating the dynamics. Thus it forms a tool to distinguish Wannier-Stark MBL from non-interacting Wannier-Stark localization up to intermediate time-scales, and to quantify quantum correlations in mixed-state dynamics.
\end{abstract}

\date{\today} 
\maketitle 



\section{Introduction}
\par Over the past decade there has been a huge research interest in disordered interacting many-body systems. It was realized that these systems could host a many-body localized (MBL) phase provided that the disorder is strong enough~\cite{Basko2006,Vosk2013,Nandkishore2015,Abanin2019}. This MBL phase is robustly non-thermalizing, and should be contrasted to a single-particle Anderson localized phase occurring in non-interacting systems~\cite{Anderson1958}. While both phases are characterized by the absence of transport, there are also notable differences, most prominently in the entanglement dynamics under a quench. In the Anderson localized system, quantum correlations cannot propagate through the system. Hence the entanglement saturates after a fast ballistic initial growth resulting from local rearrangements of particles. In the interacting MBL system, on the contrary, quantum correlations can propagate in the system but that happens only logarithmically in time~\cite{Zinidarifmmodeheckclseci2008,Bardarson2012,Serbyn2013a}. This slow growth can be understood in terms of effective exponentially decaying interactions between so-called `localized integrals of motion' (LIOMs), that form a phenomenological picture to understand MBL~\cite{Serbyn2013,Huse2014}. 
\par The existence of MBL has been experimentally confirmed~\cite{Schreiber842, Smith2015, Lueschen2017, Bordia2017}, and the logarithmic spread of quantum correlations has been observed as well~\cite{Brydges2019, Rispoli2019, Lukin2019, Chiaro2019}. Recently, there have been many proposals to establish \textit{disorder-free} types of localization~\cite{Carleo2012,DeRoeck2014,Grover2014,Schiulaz2014,Yao2014,Hickey2016,Horssen2015,Brenes2018,Smith2017,
Schulz2018,Nieuwenburg2018}, including for instance lattice gauge theories~\cite{Brenes2018}, or mixtures of two types of particles where the light ones are localized on the heavy ones~\cite{Grover2014,Schiulaz2014,Yao2014,Oppong2020,Gadway2011}. In particular, it has been realized recently that many features of MBL are also inherited by interacting systems subjected to a strong linear potential~\cite{Tomadin2007,Schulz2018,Nieuwenburg2018,Taylor2019,Yao2020,Yao2021}, yet also differences have been identified and understood in terms on Hilbert-space fragmentation~\cite{Khemani2020,Doggen2021}. In the non-interacting case this phenomenon is referred to as Wannier-Stark localization~\cite{Wannier1960,Wannier1962}; in the interacting case it is referenced to as Wannier-Stark MBL (or shortly Stark MBL in the literature). This type of localization has the advantage that it can potentially be induced solely by the tuning of an external electric field, without the need of engineering internal properties in the system. Experimental signatures of non-ergodic dynamics in systems subjected to a tilted field have been observed in Refs.~\cite{Scherg2020,Guo2020,Morong2021}, while Ref.~\cite{Guardado-Sanchez2020} was the first experiment that investigated the effect of a tilted potential on the approach to equilibrium.
\par Typical experiments are never fully isolated from the surrounding environment. This implies that it is hard to distinguish interacting types of localization from non-interacting types of localization. The most prominent difference between the two types is the logarithmic spreading of quantum correlations in the interacting (and isolated) case after a quench. In this work we therefore focus on the entanglement dynamics of interacting and non-interacting Wannier-Stark systems that are coupled to a dephasing Markovian environment. This extends our previous work~\cite{Wybo2020} where we considered a similar setup for disorder-induced MBL. Related setups have also been considered to mostly investigate transport properties in Refs.~\cite{Carmele2015,Fischer2016,Levi2016,Medvedyeva2016,Everest2017,Vakulchyk2018,Wybo2020} in the context of disorder-induced MBL, and in Ref.~\cite{Wu2019a} in the context of Wannier-Stark MBL coupled to a dephasing environment.
\par Measuring quantum correlations in open systems is challenging both theoretically as well as experimentally. From the theoretical side, it is hard to find generically computable measures that do not rely on a full diagonalization of the (partially transposed) density matrix. We circumvent this problem by considering moments of the partially transposed density matrix and calculate the third R\'{e}nyi negativity $R_3$~\cite{Calabrese2012,Calabrese2013,Gray2018,Wu2019,Wybo2020,Elben2020}, which is not an exact entanglement monotone but which captures the relevant dynamics as it behaves quantitatively similar as the negativity~\cite{Gray2018,Wybo2020}. From the experimental side, it is challenging to measure non-local correlations as full-state tomography is exponentially expensive in the system size, and as joint measurements on multiple copies of the state are also hard to engineer~\cite{Cai2008,Carteret2005,Mintert2005,Daley2012,Gray2018}. Recently there has been a lot of progress in measuring R\'{e}nyi entropies in synthetic quantum matter by random unitary measurements~\cite{Enk2012, Nakata2017, Elben2018a, Vermersch2018, Brydges2019, Elben2019a}, and this toolbox naturally includes the measurement of mixed state entanglement via R\'{e}nyi negativities~\cite{Elben2020}.

\section{Model and setup}\label{sec:model}
\par We consider the XXZ Hamiltonian with an on-site potential
\begin{equation}\label{eq:xxz}
H = J [  \sum_{i=0}^{L-2} \left( S^{x}_{i}S^{x}_{i+1} + S^{y}_{i}S^{y}_{i+1} 
+ \Delta  S^{z}_{i}S^{z}_{i+1}  \right)]  + \sum_{i=0}^{L-1} h_i S^{z}_{i} ,
\end{equation}
where $S^{x,y,z}_i$ are the spin-$\frac{1}{2}$ operators. As on-site potential we take a linear potential with small quadratic corrections of the form 
\begin{equation} \label{eq:onsite_pot}
h_i=-\gamma i + \alpha i^2/(L-1)^2 
\end{equation} 
to induce Wannier-Stark localization. Here $\gamma$ gives the slope of the linear potential, and $\alpha$ describes small quadratic corrections to it. 
\par The parameter $\Delta$ describes the strength of the nearest-neighbor interactions which can be seen by writing the Hamiltonian~\eqref{eq:xxz} in terms of spinless fermions by means of the Jordan-Wigner transformation. When there are no interactions present, $\Delta = 0$, the system is `single-particle' Wannier-Stark localized. In the presence of interactions, it is Wannier-Stark many-body localized. We will focus on the weakly interacting regime where $\Delta J < \gamma$ as to avoid resonance effects where e.g. a particle could move up the linear potential without energy cost. 
\par Localization also occurs in the same model~\eqref{eq:xxz} for randomly disordered on-site potentials. In the interacting case, there is an MBL phase if the disorder is strong enough, i.e. higher than a critical disorder strength under which the system is thermalizing.
\par The Wannier-Stark model exhibits many similar properties as the model with disorder, e.g. most importantly a slow logarithmic growth of entanglement under a quench, if the linear field $\gamma$ is sufficiently strong, and if there is some non-uniformity to this linearity~\cite{Schulz2018,Nieuwenburg2018,Taylor2019,Yao2021}. When only a linear field is applied, the (non-interacting) system contains a lot of degeneracies, and therefore properties like the level spacing statistics, can deviate from the ones of the disorder-induced localized systems. By adding a quadratic gradient most of these degeneracies are resolved, and the level-spacing statistics of the two cases are very similar~\cite{Schulz2018}.

\par
\begin{figure}
\centering
\includegraphics[width=0.4\textwidth]{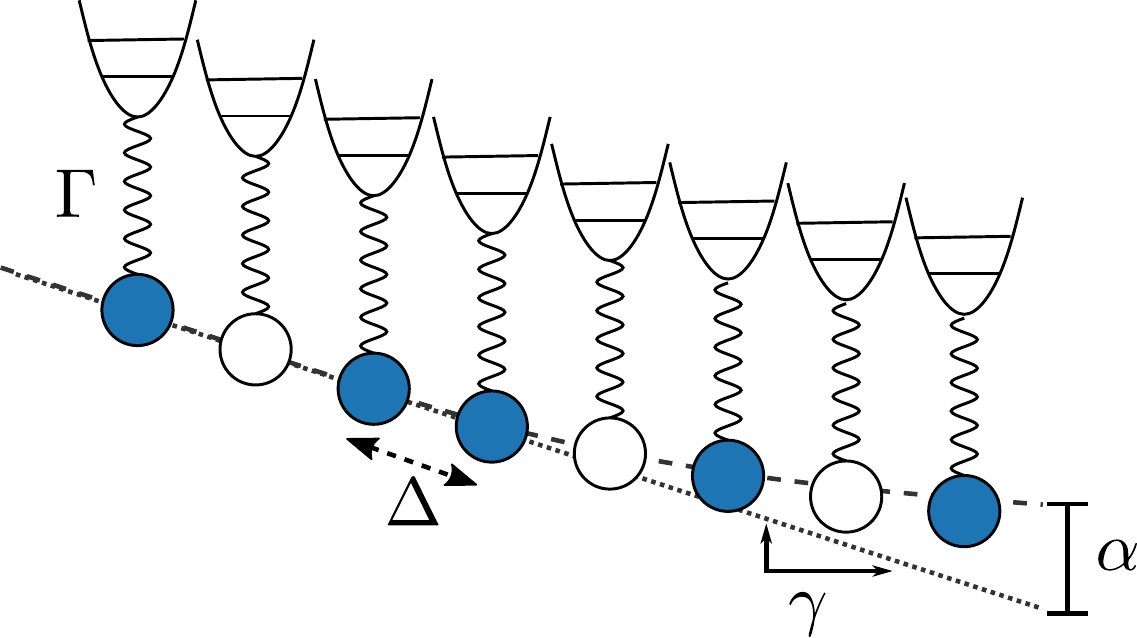}
\caption{A sketch of our setup: A chain of spins (or, equivalently spinless fermions) is subjected to a linear potential with small quadratic corrections, and is uniformly coupled to an environment.} 
\label{fig:sketch}
\end{figure}

 We couple the system to a simple, yet realistic~\cite{Lueschen2017}, Markovian dephasing environment which is modeled by the jump operators $L_i = \sqrt{\Gamma}S_i^z$, such that the time-evolution of the state is described by the following Lindblad master equation~\cite{Breuer2007}
\begin{equation}
\label{eq:lb}
\dot{\rho} = -i\left[ H, \rho(t) \right] + \Gamma \sum_i \left( 
S_i^z \rho(t) S^{z}_{i} -  \frac{1}{4} \rho(t) \right) . 
\end{equation}
Note that this type of dynamics also conserves the $U(1)$ symmetry of the Hamiltonian leading to a fixed total magnetization $M=\sum_i S_i^z$. In the limit of a purely dephasive coupling, the off-diagonal matrix elements of the density matrix decay exponentially with decay rate $\Gamma$. Hence it is clear that this type of environment coupling destroys all quantum correlations as it drives the system into a trivial infinite temperature state, that is the unique steady state of Equation~\eqref{eq:lb}. A sketch of our setup is shown in Figure~\ref{fig:sketch}. 
\par For simulating the time evolution according to the Lindblad master Equation~\eqref{eq:lb}, we use the time-evolving block decimation (TEBD) algorithm on density matrices~\cite{Verstraete2004,Zwolak2004} where the (non-unitary) time-evolution operator is given by $U(t) = \exp(\mathcal{L}t)$ with $\mathcal{L}$ the Lindblad superoperator
\begin{equation}
\mathcal{L} = -i H \otimes I + i I \otimes H + \Gamma \sum_i \left(S_i^z \otimes S^{z}_{i} -  \frac{1}{4} I \otimes I \right).
\end{equation}
This time-evolution operator acts on a (vectorized version) of the density matrix $\ket{\rho}$ as $\ket{\rho(t+\dd t}=$ \\ $\exp(\mathcal{L}\dd t)\ket{\rho(t)}$. The matrix-product decomposition of $\ket{\rho}$ reads
\begin{equation}
\ket{\rho} = \sum_{\substack{i_1,\dots, i_L \\ j_1,\dots, j_L }}^{d=2} M_{[1]}^{i_1,j_1} M_{[2]}^{i_2,j_2} \dots M_{[L]}^{i_L,j_L} \ket{i_1j_1,\dots, i_Lj_L},
\end{equation}
where the matrices $M_{[k]}^{i_k,j_k}$ have dimension $\chi_{k-1}\times \chi_{k}$ where $\max(\chi_{k-1},\chi_{k}) \leq \chi \leq d^L$, with $\chi$ the maximal bond dimension we allow for in the simulation. Hence the so-called operator entanglement entropy of  $\ket{\rho} $ is limited to $\log\chi$ during the evolution towards the steady state. Because of the localized nature of the isolated system and the addition of dephasing noise, the operator entanglement entropy stays moderate at all times in the considered setups.
For more details about the numerical implementation, we refer to our previous work Ref.~\cite{Wybo2020}. 

\begin{figure}
\centering
 \includegraphics[width=0.46\textwidth]{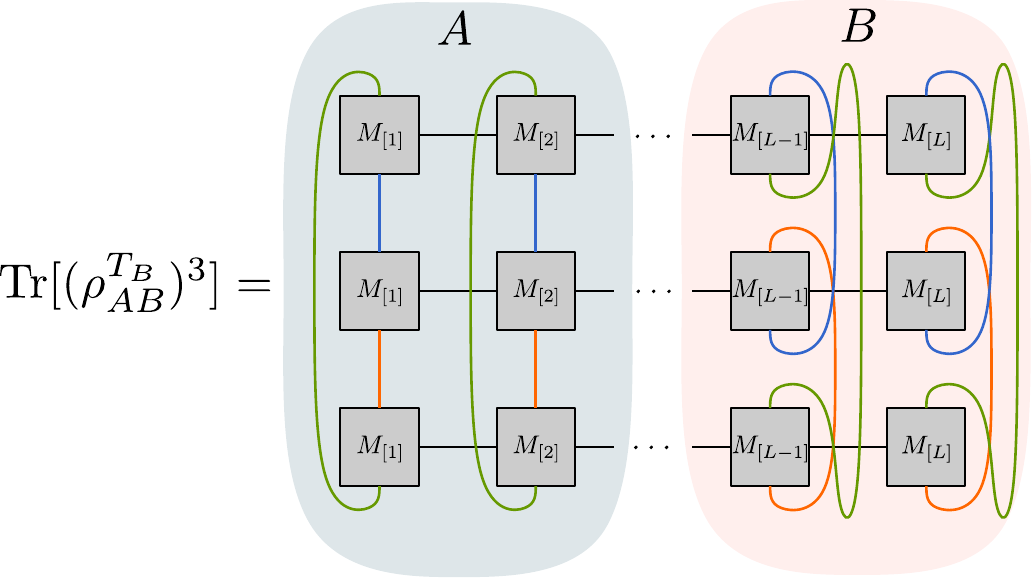}
\caption{A sketch of the contraction of the three layer tensor network to compute $\Tr [(\rho_{AB}^{T_B})^3]$. We bipartition the system into two equal subsystems $A$ and $B$, and take the transpose of the degrees of freedom of partition $B$. After which we take the trace of the system (green lines). }
\label{fig:sketch_r3}
\end{figure}

\par We study a global quench setup by starting from an initial (pure) product state $\rho_0 = \ket{\psi_0}\bra{\psi_0}$ where $\ket{\psi_0}$ is a product state in the zero magnetization sector. Our goal is to measure the dynamics quantum of quantum correlations in such a setup. However, in a mixed state many-body context it is very challenging to separate quantum correlations from classical correlations in a generic way. There exist experimentally accessible quantities like the quantum Fisher information~\cite{Braunstein1994,Hyllus2012, Toth2012, Smith2015}, however these so-called entanglement witnesses rely on system specific properties and are hence not as generic. 
\par A famous generic criterion is the so-called positive partial transpose (PPT) criterion of Peres~\cite{Peres1996}. It states that if a bipartite density matrix $\rho_{AB}$ is separable, then $\rho_{AB}^{T_B}$ has only positive eigenvalues, where ${T_B}$ denotes the operation of partial transposition. The negativity is then a generic entanglement monotone that quantifies the violation of this criterion~\cite{Vidal2002} 
\begin{equation}
\mathcal{N}(\rho_{AB}) = \frac{\|\rho_{AB}^{T_B}\|_1-1}{2},
\end{equation}
however, because of the trace norm, this quantity relies on the full diagonalization of $\rho_{AB}^{T_B}$ which is exponentially hard in the system size. Therefore the negativity is not computable for large systems.
\par However, as detailed in Ref.~\cite{Gray2018,Wu2019, Wybo2020} we can use the third R\'{e}nyi negativity to estimate the amount of bipartite entanglement in the system 
\begin{equation} \label{eq:R3}
R_3(\rho_{AB}) = -\log \left(  \dfrac{\Tr [(\rho_{AB}^{T_B})^3]}{ \Tr\rho_{AB}^3}\right).
\end{equation}
Note that the equivalent quantities using the first and second power of the density matrix are trivially zero as $\Tr \rho^{T_B}_{AB} = \Tr \rho_{AB} = 1$ and $\Tr (\rho_{AB}^{T_B})^2 = \Tr\rho_{AB}^2$.
The quantity $R_3$ is however not an entanglement monotone as the negativity (for the definition of an entanglement monotone see e.g. Ref.~\cite{Guehne2008}) as it is always zero for simple product states, but not necessarily for separable (`classically' correlated) states. However, in case of disordered MBL we have shown in Ref.~\cite{Wybo2020} that it has the same dynamical properties as the negativity. This is manifested for instance by very similar stretching exponents. The contraction of the network to compute $\Tr[ (\rho_{AB}^{T_B})^3]$ is sketched in Figure~\ref{fig:sketch_r3}.  
\par For a pure state, the density matrix has the form $\rho_{AB} = \ket{\psi_{AB}}\bra{\psi_{AB}}$, hence we have that
\begin{equation} 
R_3(\rho_{AB})= -\log \Tr \rho_B^3 = 2 S_3(\psi_{AB}),
\end{equation}
and thus reduces to the third R\'{e}nyi entropy~\cite{Calabrese2013}. Here, $\rho_B = \Tr_A \rho_{AB}$ denotes the reduced density matrix of subsystem $B$. 
\par In what follows we will always assume that we have two equal partitions in the system and shortly denote $R_3(\rho_{AB}) \equiv R_3$.

\section{Results}
In this section, we discuss our results. In the closed case, we first look at the entanglement dynamics by making a quench from a N\'{e}el state, and secondly at the dynamics by making a quench from a random product state in the zero magnetization sector and average over the results. Afterwards, we discuss the results in the open case $\Gamma >0$.

\subsection{Isolated system}
We start by investigating the dynamics of $R_3$ in the isolated system $\Gamma = 0$. In this case we have that $R_3$ is directly related to the third R\'{e}nyi entropy as discussed in the previous section. 

\subsubsection{Quench from a N\'{e}el state}

\begin{figure}
\centering
\includegraphics[width=0.48\textwidth]{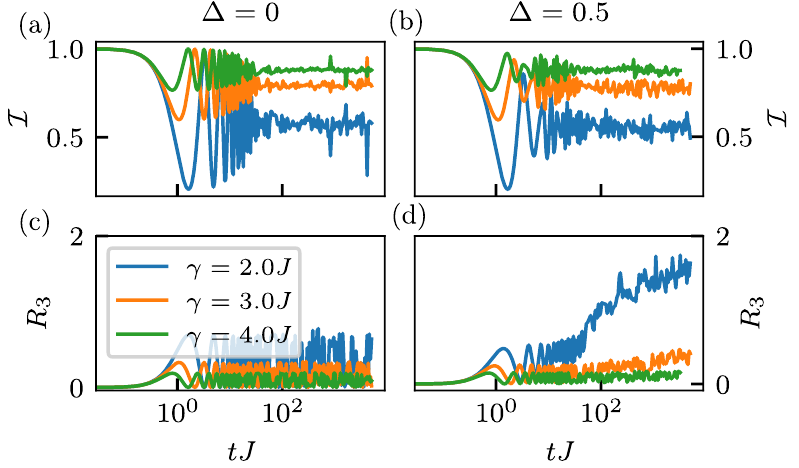}
\caption{The evolution of the imbalance [(a)-(b)] and third R\'{e}nyi negativity [(c)-(d)] in closed Wannier-Stark localized systems of $L=24$ sites. The slope of the linear potential $\gamma$ is given in the legend and the quadratic corrections have a strength of $\alpha = 2J$. The most striking difference between the non-interacting localized case (c) and the interacting localized case (d) is that entanglement can propagate through the system in the interacting case, most notably when the slope of the linear potential is not too strong.} 
\label{fig:Neel_isolated}
\end{figure}

When we start initially from a N\'{e}el state $\ket{\downarrow \uparrow \downarrow \uparrow  \ldots }$, we can track how fast the Wannier-Stark MBL system `loses' information about this initial state pattern. This can be realized by considering a quantity like the imbalance
\begin{equation}\label{eq:imb}
\mathcal{I}=\frac{\expval{N_e^z-N_o^z}}{\expval{N_e^z+N_o^z}} 
\end{equation}
where $N_e^z/N_o^z$ is summing over the occupation numbers $N_i^z=S_i^z +1/2$ for even/odd sites. For localized systems $\mathcal{I}$ decays to a highly non-thermal (i.e. non-zero) value, and this forms a simple and accessible experimental probe for localization~\cite{Schreiber842}. However, a priori the imbalance decay does not directly allow one to distinguish interacting types of localization from non-interacting types of localization. This is shown in Figure~\ref{fig:Neel_isolated}, where we show both the imbalance and entanglement dynamics in the interacting ($\Delta = 0.5$) and non-interacting ($\Delta = 0$) cases for various slopes of the linear potential. From this it is clear that only the entanglement dynamics provides a striking difference between the two cases.

\subsubsection{Quench from a random initial state}
\par
\begin{figure}
\centering
\includegraphics[width=0.4\textwidth]{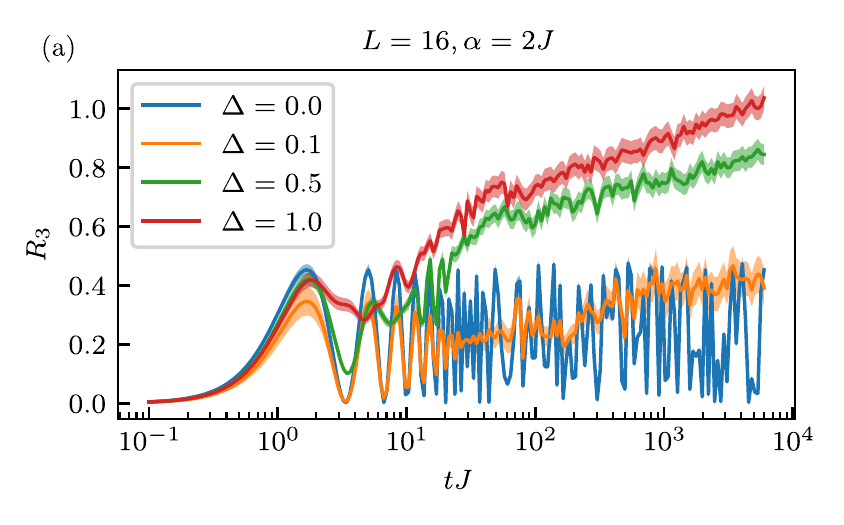}
\includegraphics[width=0.4\textwidth]{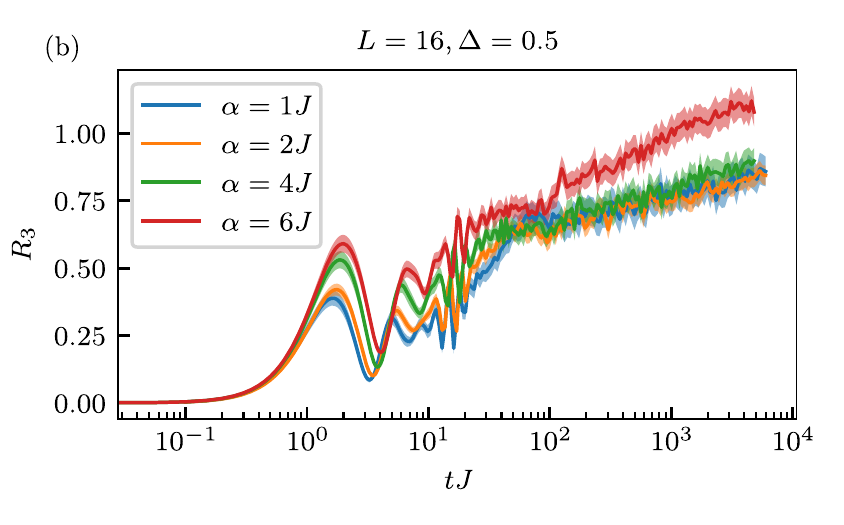}
\includegraphics[width=0.4\textwidth]{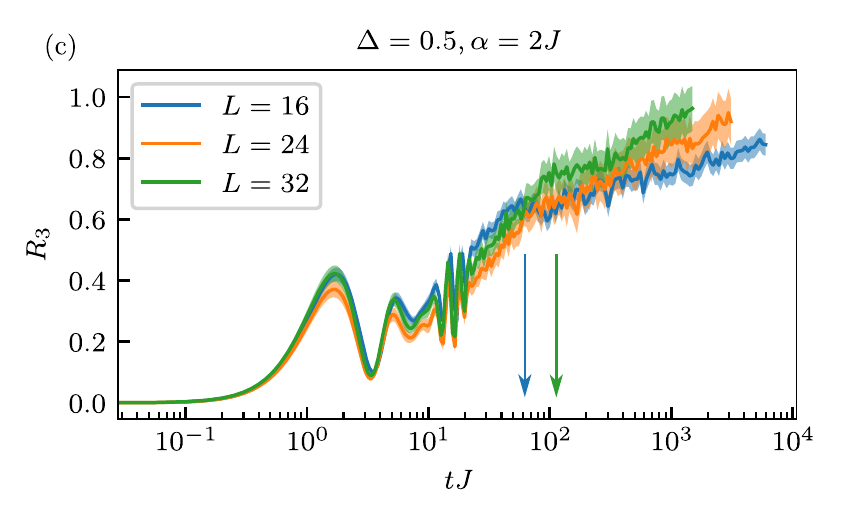}
\caption{The growth of entanglement quantified by the third R\'{e}nyi negativity (entropy) $R_3$ in the isolated system. (a) When the system size is fixed, the interactions determine the onset of logarithmic growth. (b) When the potential parameter $\alpha$ is decreased the finite-time cross-over becomes more pronounced. (c) If only the system size is varied, $t_{\textrm{cross}}$ is doubled with system sizes as indicated by the arrows.  } 
\label{fig:closed}
\end{figure}

 We now consider a random product state in the zero magnetization sector and average the results over the different realizations. This can be seen in Figure~\ref{fig:closed}(a), where we show the logarithmic growth of $R_3$ for different interaction strengths at field parameters $\alpha = \gamma =2J$. Without interactions $\Delta = 0$ the system is single-particle Wannier-Stark localized and entanglement cannot propagate through the system.  Fluctuations are stronger in the non-interacting limit, as the level spacing is only inversely proportional to the system size. The time-scale at which the effect of the interactions becomes dominant is set by $t_{\textrm{int}} \sim (\Delta J)^{-1}$. In the data obtained for finite interactions, a cross-over time-scale $t_{\textrm{cross}}$ becomes apparent, which is absent in the case of disorder-induced MBL, where there is a faster logarithmic growth, after the initial ballistic growth up to times $tJ\sim 1$. Beyond this cross-over regime, there is then a slower, logarithmic growth. We attribute the existence of this cross-over regime to the quadratic contribution of the potential of strength $\alpha$ in~\eqref{eq:onsite_pot}. When increasing the quadratic deviations, the cross-over regime becomes less pronounced, as we show in Figure~\ref{fig:closed}(b). 
\par In Figure~\ref{fig:closed}(c) we show the logarithmic growth at one particular interaction strength $\Delta = 0.5$ for various system sizes. Also, $t_{\textrm{cross}}$ is doubled when the system size is doubled which confirms that the parameter $\alpha$ of our local potential~\eqref{eq:onsite_pot} is indeed governing $t_{\textrm{cross}}$. When increasing system size, $\alpha$ thus has to be rescaled accordingly when considering the in the literature commonly used form of the local potential~\cite{Schulz2018}.   

\subsection{Open System}
\par
\begin{figure}
\centering
\includegraphics[width=0.48\textwidth]{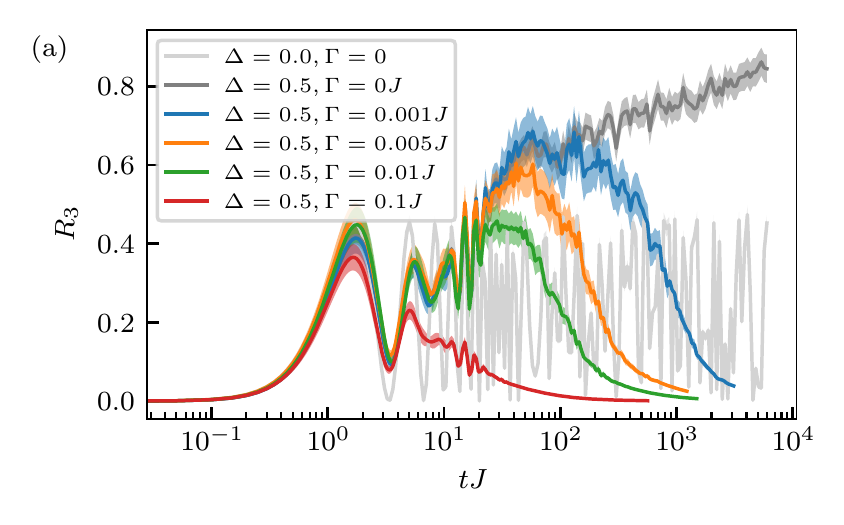} 
\includegraphics[width=0.48\textwidth]{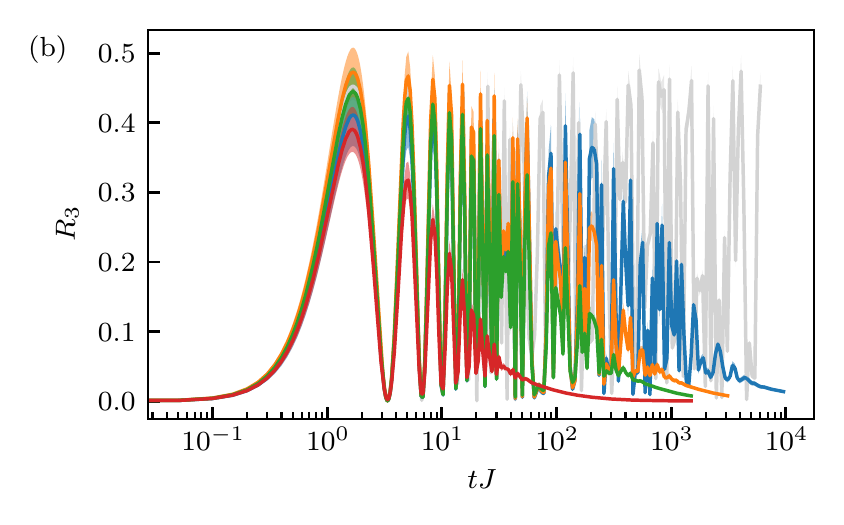} 
\caption{The dynamics of $R_3$ in a Wannier-Stark system of length $L=16$ with $\alpha=2J$, $\gamma=2J$ coupled to a dephasive environment. (a) Interacting case $\Delta = 0.5$, where we show the data for the closed non-interacting case for comparison (light grey line). $R_3$ reproduces the onset of entanglement growth for sufficiently weak dephasing. (b) Non-interacting case $\Delta=0$. The coupling strengths are given in the legend. We averaged over about $100$ initial states in the $M=0$ sector for $\Gamma >0$, and over about $300$ states for $\Gamma=0$ (exact diagonalization was used in this case).   }
\label{fig:av_L16}
\end{figure}

 We now turn to the investigation of the dynamics of $R_3$ in the open system $\Gamma > 0$. When the system is coupled to a dephasing enviroment, all entanglement structure will be eventually lost as the system heats up to the infinite temperature state. The time scale on which the depasing starts to dominate the dynamics is set by the coupling strength $t_{\textrm{deph}} \sim 1/\Gamma$. In order to allow that the interactions can still dominate the dynamics at intermediate times, we need to have that $t_{\textrm{int}}  \ll t \ll t_{\textrm{deph}}$ which implies that we must have that $\Gamma/J \ll \Delta $. Hence, the dephasing strength must be sufficiently weak compared to the interaction strength to be able to still observe signatures of the logarithmic entanglement growth. In Figure~\ref{fig:av_L16} we show the entanglement dynamics for various coupling strengths. One of the main advantages of looking at the R\'{e}nyi negativity $R_3$ is that we can capture the logarithmic growth of quantum correlations even if the system becomes slightly mixed. This onset of logarithmic growth can indeed be still observed for sufficiently weak dephasing strengths in Figure~\ref{fig:av_L16}(a). In principle, for characterizing interacting dynamics it is sufficient that the entanglement reaches a value that is higher than the maximal value of the oscillations in the non-interacting case.  The decay of $R_3$ in the non-interacting case is shown in Figure~\ref{fig:av_L16}(b) for comparison.

\begin{figure}[t!]
\centering
\includegraphics[width=0.48\textwidth]{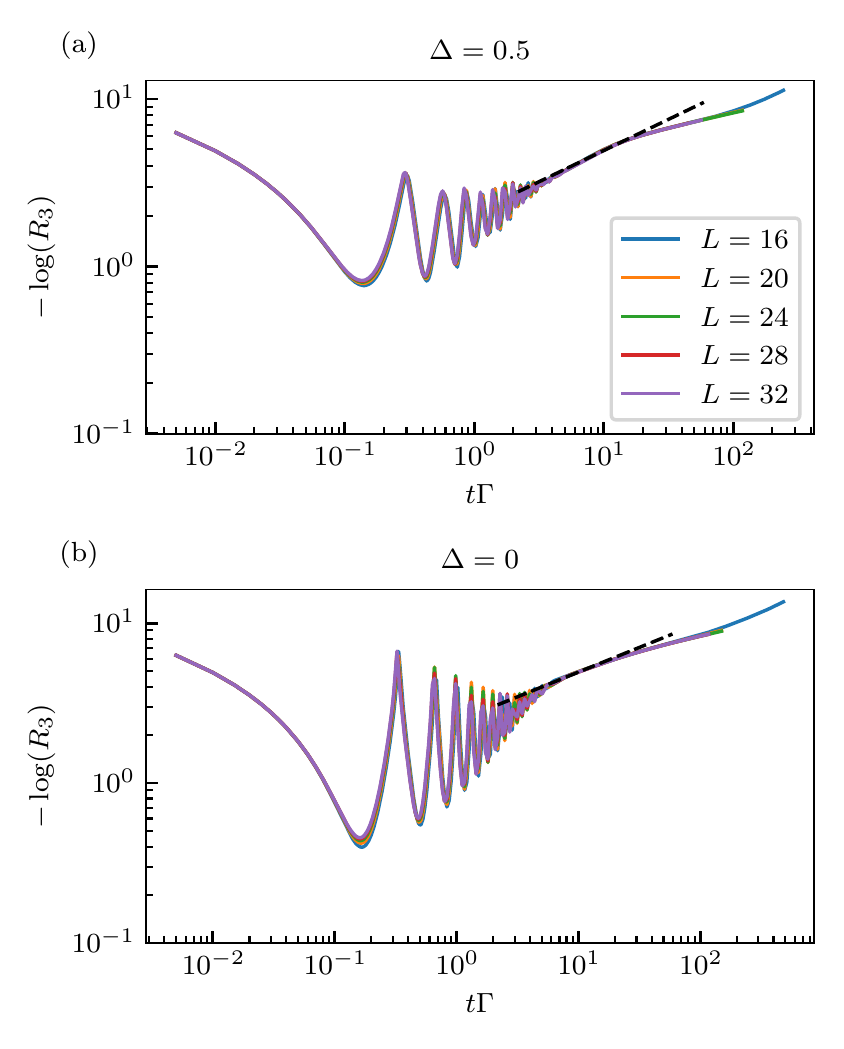} 
\caption{The dynamics of $R_3$ in a Wannier-Stark system with $\alpha=2J$, $\gamma=2J$ coupled to a dephasive environment under a quench from the N\'{e}el state. The black dashed lines are stretched-exponential fits to intermediate-time regime of $L=16$. (a) Interacting case $\Delta = 0.5$, the slope of the fit shown is $b \approx 0.37$. (b) Non-interacting case $\Delta=0$, the slope of the fit shown is $b \approx 0.31$. In both cases the late-time dynamics deviates from stretched exponential decay. } 
\label{fig:neel_r3}
\end{figure} 

\begin{figure}[t!]
\centering
\includegraphics[width=0.48\textwidth]{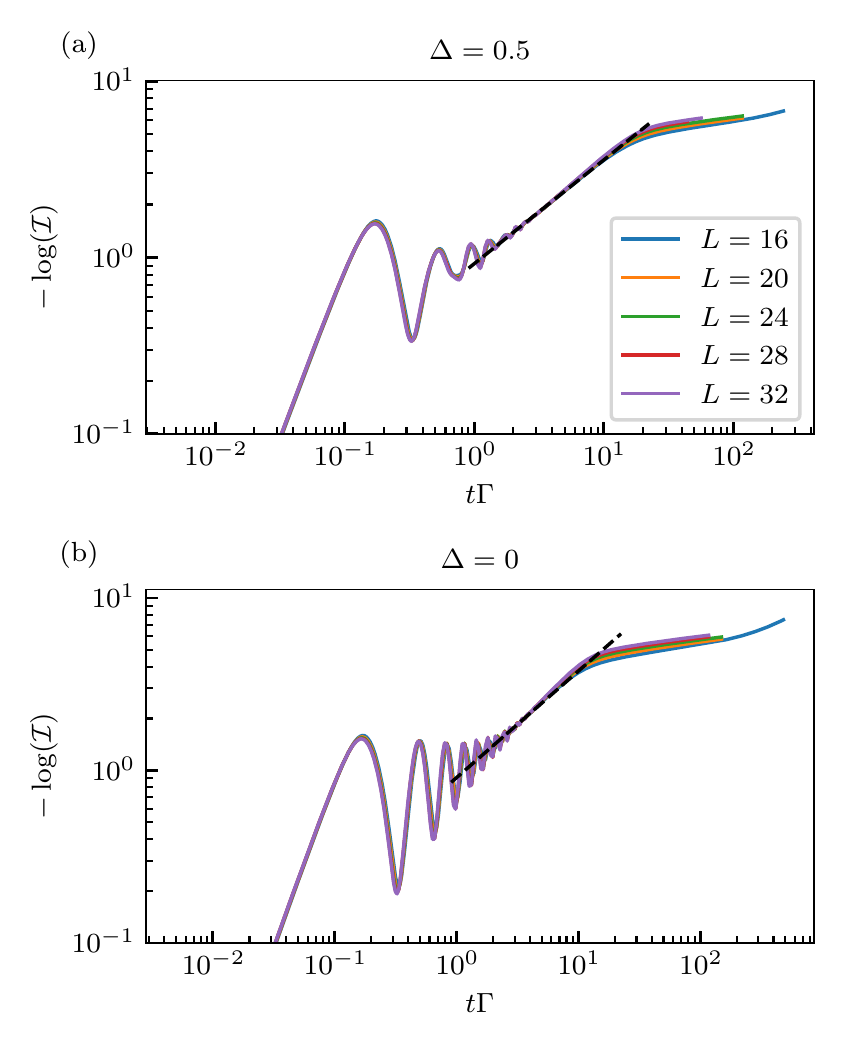} 
\caption{The dynamics of the imbalance in a Wannier-Stark system with $\alpha=2J$, $\gamma=2J$ coupled to a dephasive environment under a quench from the N\'{e}el state. The black dashed lines are stretched-exponential fits for $L=16$. (a) Interacting case $\Delta = 0.5$, the slope of the fit shown is $b \approx 0.59$. (b) Non-interacting case $\Delta=0$, the slope of the fit shown is $b \approx 0.62$. 
Again, in both cases the late-time dynamics is incompatible with stretched exponential decay. }
\label{fig:neel_imb}
\end{figure} 

\par From disordered MBL it is known that the tails of the decay of the imbalance~\cite{Fischer2016,Levi2016} and the negativity or $R_3$~\cite{Wybo2020} are stretched exponentials $\sim e^{-({\Gamma t/a})^b}$ with $b<1$. These stretched exponentials are in that case understood as a superposition of many local exponential decays~\cite{Fischer2016}, originating from very broad distributions of exponentially decaying couplings in the phenomenological LIOM picture of MBL~\cite{Varma2019,Serbyn2013,Huse2014}. However, the Wannier-Stark case is disorder free, which could therefore lead to a different behavior of the tails of the decay. In the non-interacting case, such a difference has been reported in the decay of the imbalance in Ref.~\cite{Wu2019a}. 
\par In Figures~\ref{fig:neel_r3} and ~\ref{fig:neel_imb}, we show the decay of respectively $R_3$ and the imbalance under a quench from the N\'{e}el state with $\Gamma = 0.1$. These figures have a double logarithmic scale on the $y$-axis and single logarithmic scale on the $x$-axis, such that a stretched exponential $\sim e^{-({\Gamma t/a})^b}$ would appear as a straight line with $b$ the slope, and $a$ related to the offset. We have fitted stretched exponentials to the data in the intermediate-time regime. However, in the late-time regime, the functional form of the decay changes. For small system sizes, exponential decay can be observed at late times as we illustrate in the Appendix~\ref{app:tails}. In this Appendix we also demonstrate that averaging over initial states does not lead to a quantitatively different behavior in the decay of $R_3$.
\par In Ref.~\cite{Wu2019a} they report that the decay of the imbalance happens according to an exponential in the non-interacting case, while in the interacting case it happens according to a stretched exponential. They however consider much stronger tilts and interactions. As we focus here on a more moderate regime of weaker tilts and weak interactions as relevant for experiments~\cite{Taylor2019}, we can not distinguish qualitative differences between both cases in the late-time dynamics.

\section{Conclusion}
\par In this work we have investigated the dynamics of entanglement in a non-interacting Wannier-Stark localized and in an interacting Wannier-Stark many-body localized system. In the closed case, we have observed a cross-over regime that is absent in the case of disordered MBL and that is related to the quadratic corrections from linearity of the field. In the open system with dephasing noise, it is possible to still observe parts of the characteristic logarithmic growth for sufficiently weak dephasing strengths. 
\par Our results confirm that the Wannier-Stark MBL system indeed inherits many properties from the disordered MBL system in particular when considering the entanglement dynamics coupled to an environment. However, we do not observe a robust stretched-exponential functional form of the entanglement decay in the Wannier-Stark case.
\par For future work, it would be interesting to investigate the entanglement dynamics under different types of dissipation. In particular for non-hermitean Lindblad operators, it could be interesting to investigate whether a physically relevant entanglement structure could still exist in the non-trivial steady state.

\medskip
\textbf{Acknowledgements} \par 
Our tensor-network calculations were performed using the TeNPy Library~\cite{Hauschild2018}.
We acknowledge support from the Technical University of Munich - Institute for Advanced Study, funded by the German Excellence Initiative and the European Union FP7 under grant agreement 291763, the Deutsche Forschungsgemeinschaft (DFG, German Research Foundation) under Germanys Excellence Strategy-EXC-2111-390814868, DFG TRR80 project number 107745057 and DFG grant No. KN1254/1-1, and from the European Research Council (ERC) under the European Unions Horizon 2020 research and innovation programme (grant agreements No. 771537 and 851161).

\medskip
%
\bibliography{biblio}

\appendix
\section{Late-time dynamics} \label{app:tails}
\par
\begin{figure}
\centering
\includegraphics[width=0.48\textwidth]{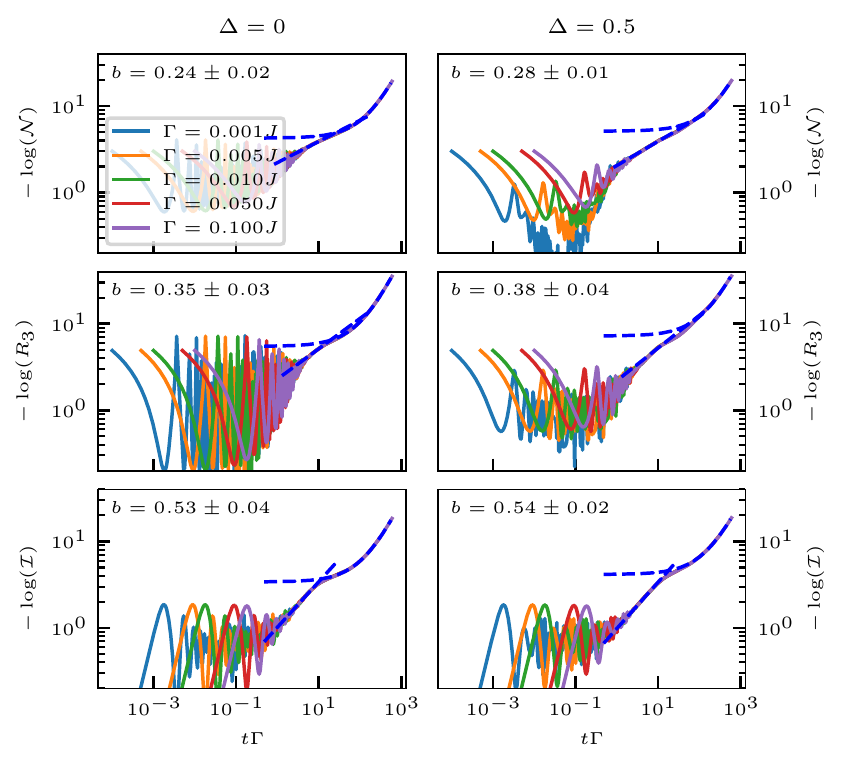}
\caption{The dynamics of the negativity, $R_3$, and the imbalance for a system of $L=8$ sites under a quench starting from the N\'{e}el state. The tilted potential is characterized by $\alpha=\gamma=2J$. The non-interacting ($\Delta =0$) and interacting ($\Delta =0.5$) cases are shown. The blue dashed lines show a stretched-exponential fit $\sim e^{-({\Gamma t/a})^b}$ for the intermediate-time regime and an exponential fit for the late-time regime. The stretching exponent $b$ is written inside the panels with $3\sigma$ confidence level.}
\label{fig:neel_ed}
\end{figure}

\begin{figure}
\centering
\includegraphics[width=0.48\textwidth]{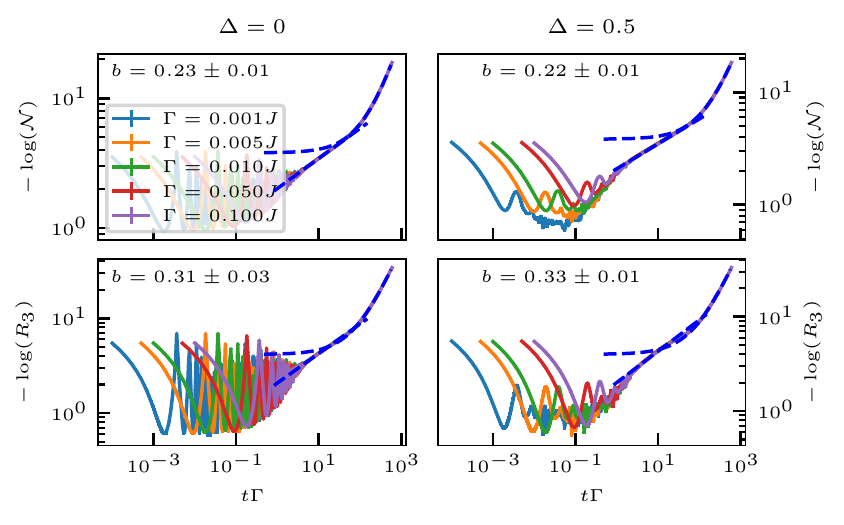}
\caption{The dynamics of the negativity and $R_3$ for a system of $L=8$ sites under a quench starting from a random initial state. Averages are taken over initial product states in the $M=0$ sector. The tilted potential is characterized by $\alpha=\gamma=2J$. The non-interacting ($\Delta =0$) and interacting ($\Delta =0.5$) cases are shown. The blue dashed lines show a stretched-exponential fit $\sim e^{-({\Gamma t/a})^b}$ for the intermediate-time regime and an exponential fit for the late-time regime. The stretching exponent $b$ is written inside the panels with $3\sigma$ confidence level. }
\label{fig:av_ed}
\end{figure}

\begin{figure}
\centering
\includegraphics[width=0.48\textwidth]{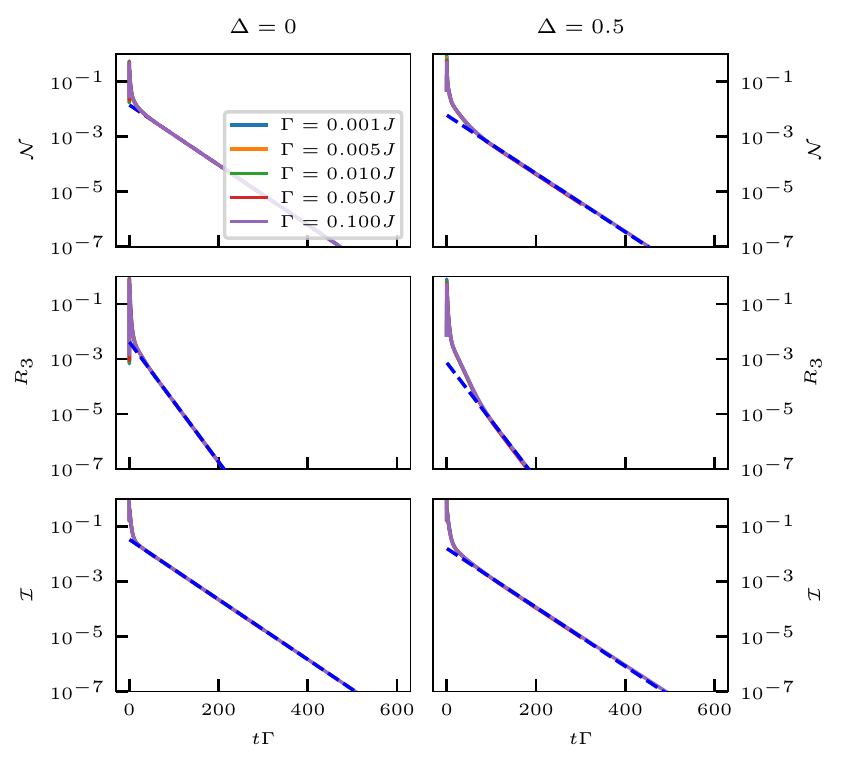}
\caption{The same data as in Figure~\ref{fig:neel_ed} but in a different scale to make the exponential form of the tails visible.}
\label{fig:neel_ed_exp}
\end{figure}

\begin{figure}
\centering
\includegraphics[width=0.48\textwidth]{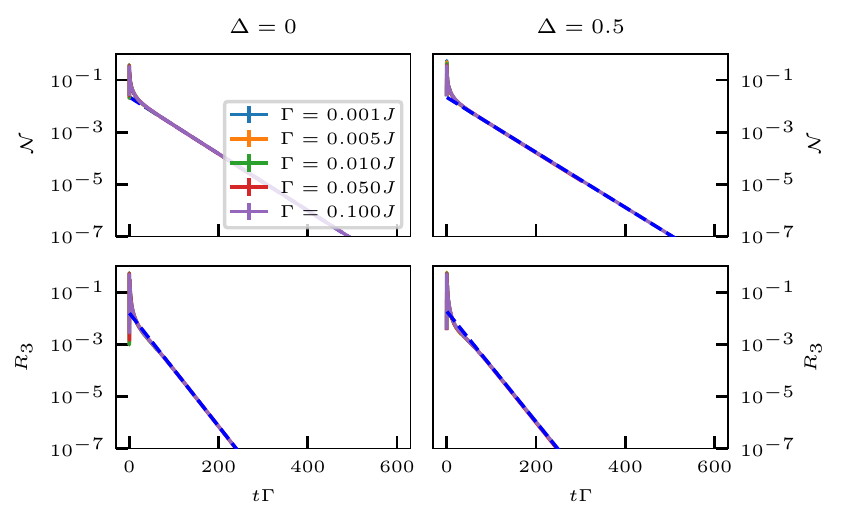}
\caption{The same data as in Figure~\ref{fig:av_ed} but in a different scale to make the exponential form of the tails visible.}
\label{fig:av_ed_exp}
\end{figure}

 In this Appendix, we present some exact diagonalization data, obtained by numerically integrating the Lindblad Equation~\eqref{eq:lb} for a small system of $L=8$ sites. In Figure~\ref{fig:neel_ed} we show the collapse of the tails for various coupling strengths $\Gamma$ for a quench starting from the N\'{e}el state. We have fitted stretched exponentials $\sim e^{-({\Gamma t/a})^b}$ in the intermediate-time regime, and exponentials $\sim \alpha e^{-{\Gamma t/\beta}}$ in the late-time regime. In Figure~\ref{fig:neel_ed_exp} we show the same data in a different scale that makes the exponential form of the late-time tails clearly visible. In Figures~\ref{fig:av_ed} and~\ref{fig:av_ed_exp}, we show the data when we average over different initial states. From this, it is clear that the averaging does not alter the behavior of the tails quantitatively, e.g. the best-fit stretching exponents in the intermediate-time regime stay very similar as can be seen by comparing Figures~\ref{fig:neel_ed} and~\ref{fig:av_ed}.

\end{document}